# Long-lasting activity of asteroid (248370) 2005 QN$_{173}$

Oleksandra Ivanova, [1,2,3]★ Javier Licandro, [4,5]★ Fernando Moreno, [6]★ Igor Luk'yanyk, [3] Johannes Markkanen, [7,8] Dušan Tomko, [1] Marek Husárik, [1] Antonio Cabrera-Lavers, [9,4,5] Marcel Popescu, [10,11] Elena Shablovinskaya[12] and Olena Shubina[1,2]

[1] *Astronomical Institute of the Slovak Academy of Sciences, 059 60 Tatranská Lomnica, Slovak Republic, Slovak Republic*
[2] *Main Astronomical Observatory of National Academy of Sciences of Ukraine, Kyiv 03143, Ukraine*
[3] *Astronomical Observatory of Taras Shevchenko National University of Kyiv, Kyiv, 04053, Ukraine*
[4] *Instituto de Astrofisica de Canarias, Research, La Laguna, Tenerife 38205, Spain*
[5] *Departamento de Astrofsica, Universidad de La Laguna – ULL, Tenerife 38200, Spain*
[6] *Instituto de Astrofísica de Andaluaá, CSIC, Glorieta de la Astronomá s/n, E-18008 Granada, Spain*
[7] *Institut für Geophysik und Extraterrestrische Physik, Technische Universität Braunschweig, Mendelssohnstr. 3, D-38106, Braunschweig, Germany*
[8] *Max Planck Institute for Solar System Research, Justus-von-Liebig-Weg 3, D-37077, Göttingen, Germany*
[9] *GRANTECAN, Cuesta de San José s/n, E-38712 Breña Baja, La Palma, Spain*
[10] *Astronomical Institute of the Romanian Academy, 5 Cuţitul de Argint, 040557 Bucharest, Romania*
[11] *A Faculty of Physics, University of Bucharest, 405 Atomiştilor str., Măgurele 077125, Ilfov, Romania*
[12] *Núcleo de Astronomá de la Facultad de Ingeniería, Universidad Diego Portales, Av. Ejército Libertador 441, Santiago, Chile*

**ABSTRACT**

We present the results of observations of asteroid (248370) QN$_{173}$ obtained during July 2021–January 2022 with three telescopes. Our analysis revealed the presence of the dust tail for about half of a year. The direct images of the asteroid were obtained with broad-band filters. No emissions were revealed in the spectra, and the spectrum of the asteroid closely matched that of a C-type asteroid. Created colour and linear polarization variations along the tail were analysed. The asteroid demonstrated a redder colour compared to the Sun. Dramatic changes in dust productivity obtained in different filters were not detected. The $g - r$ colour changes from $0.2^m$ to $0.7^m$ over the coma, and the linear polarization degree varies from about 1.2 per cent to 0.2 per cent and from $-0.2$ per cent to $-1.5$ per cent at the phase angle of $23.2°$ and $8.16°$. The total dust mass ejected until the latest observation on October 10 is $4.2 \times 10^7$ kg, with a maximum rate of 2.6 kg s$^{-1}$ based on the Monte Carlo modelling of the dust tail. The estimated asteroid size is 1.3 km. It is shown that large particles are concentrated around the nucleus, whereas smaller ones dominate in the tail. The evolution of (248370) QN$_{173}$ orbit and the orbits of the sample of the 464 short-periodic comets were followed. Ten of them approached the asteroid's orbit. These objects are not genetically related, despite very close distance of their orbits for a relatively long time.

## 1 INTRODUCTION

Active processes in asteroids, comets, or planets' satellites are often accompanied by the release of primary matter from their inner parts, the composition and physical properties of which show the conditions and evolution of these bodies' formation, which is of great cosmogonic importance. In this sense, the genetic link between comets, asteroids, and planets' satellites is crucial. Sometimes, it is difficult to determine to which class of small celestial bodies an object belongs. For example, many satellites of planets with retrograde orbits are more like captured asteroids than other planetary satellites that rotate in the plane of the equator in circular orbits (Ma, Zheng & Shen 2010). There were also discovered objects belonging to both classes: asteroids and comets (Jewitt 2012). They have the dynamic characteristics of asteroids (asteroidal orbits) and demonstrate cometary activity expressed in the appearance of dust coma and tails. The cometary activity of such asteroids may be periodic (Jewitt 2012; Chandler et al. 2019) or short-time (Jewitt et al. 2010; Snodgrass et al. 2010; Neslusan et al. 2016). Several mechanisms of mass-loss for 11 such objects are considered in papers (Jewitt 2012; Busarev et al. 2019), where they are called 'active asteroids'. The key to understanding the genetic links and the evolution of such objects is to explain the nature of their activity, the study of which has its specific features. So, observations of active asteroids are extremely rare because of unexpecting and unpredictable objects' activity. Also, in contrast with regular observations of comets, typically characterized long-term activity observations of active asteroids are often carried out during limited time due to observational conditions and relatively short duration. The formation of a comet-like coma in asteroids is a temporary process with a limited duration. Once the coma is formed, it exhibits a relatively short lifetime. Its activation may require the simultaneous fulfillment of certain conditions, for example, a high content of ice in the surface substance and an increase in surface temperature. Sublimation activity has recently been detected in the C-type asteroids (145) Adeona, (704) Interamnia, (779) Nina, and (1474) Beira based on spectral observations (Busarev, Barabanov & Puzin 2016) that indicated a significant content of water ice in their matters. Apparently, the simultaneity of this phenomenon on several asteroids of primitive types points out the similarity of their characteristics and, accordingly, the conditions of their origin.

★ E-mail: oivanova@ta3.sk (OI); jlicandr@iac.es (JL); fernando@iaa.es (FM)

**Table 1.** Orbital parameters of active asteroid (248370) 2005 QN$_{173}$ (J2000.0): eccentricity, semimajor axis, perihelion distance, inclination, orbital period, Jupiter Tisser, and invariant.

| $e$ | $a$, au | $q$, au | $i$, deg | $P$, y | $T_{Jup}$ |
|---|---|---|---|---|---|
| 0.22555 | 3.064 | 2.737 | 0.0677 | 5.36 | 3.193 |

In this work, we have focused on a comprehensive analysis of observations of an active asteroid (248370) 2005 QN$_{173}$ (hereafter 2005 QN$_{173}$). The main-belt asteroid 2005 QN$_{173}$, also designated as comet 433P, was discovered on 2005 August 29, by the NEAT survey at Palomar. Its orbital parameters taken from NASA JPL Small-Body Data base[1] are listed in Table 1. The asteroid 2005 QN$_{173}$ has previously been measured to have a diameter of 3.6 ± 0.2 km and visible geometric albedo of 0.054 ± 0.012 for the V-band absolute magnitude and $G = 0.15$ (Mainzer et al. 2019). Hsieh et al. (2021) using object's V-band albedo 0.054 ± 0.012 (Mainzer et al. 2019) and apparent V-band magnitude of the Sun −26.71 ± 0.03 (Hardorp 1980) found an effective asteroid radius 1.6 ± 0.2 km, which is slightly smaller than the radius computed by Mainzer et al. (2019). The taxonomic class of asteroid 2005 QN$_{173}$ is not yet known, but it is likely a C-type asteroid (Chandler, Trujillo & Hsieh 2021; Hsieh et al. 2021). Novaković et al. (2022) estimated two possible rotation period of 2.7 ± 0.1 and 4.1 ± 0.1 h. The activity of an asteroid was detected on 2021 July 7 in data obtained by the Asteroid Terrestrial-impact Last Alert System survey (ATLAS; Tonry et al. 2018a). Observation presented in Hsieh et al. (2021) supported the discovered activity of the asteroid. The authors estimated dust productivity $Af\rho$, which changed from 7 to 27 cm in July–August 2021. Hsieh et al. (2021) estimated the colours of $g' - r' = 0.47 \pm 0.03$, $r' - i' = 0.10 \pm 0.04$, and $i' - z' = 0.05 \pm 0.05$ for the near-nucleus coma of the asteroid and similar values for dust tail colours. It allows the authors to suggest that no significant gas coma is present. Chandler et al. (2021) based on archival observations of 2005 QN$_{173}$ found that the asteroid demonstrated at least one more apparition of activity near perihelion. Authors discovered evidence of this second activity epoch in an image captured on 2016 July 22, with the DECam on the 4-m Blanco telescope at the Cerro Tololo Inter-American Observatory in Chile. The mechanism of asteroid activity is discussed in the literature (Chandler et al. 2021; Hsieh et al. 2021) but is still an open question. Chandler et al. (2021), based on the paper by Jewitt et al. (2018), concluded there is no evidence of the binary nature of 248 370 caused by the repeatedly colliding of two merging asteroids with material ejection.

Our observations were obtained from July 2021 to January 2022. This period covers both the state of activity and the period without a pronounced coma and tail. The paper is organized as follows. The observations and data reduction technique are presented in Section 2. The main findings based on observed data are presented and discussed in Section 3. The dynamic modelling of the asteroid motion is described in Section 4. A summary of the obtained results can be found in Section 5.

## 2 OBSERVATIONS AND DATA REDUCTIONS

We observed 2005 QN$_{173}$ from 2021 July 10 to 2022 January 28 using three different telescopes: the world largest optical 10.4-m Gran Telescopio Canarias (GTC) located in the 'El Roque de los Muchachos' observatory (ORM, La Palma, Spain), 6-m Big Telescope Altazimuth (BTA) telescope of the Special Astrophysical Observatory, and the 2.5-m telescope of the Caucasian Mountain Observatory (CMO). We obtained images, spectroscopic, and polarimetric data in different sessions.

The technical information about the telescopes and CCD cameras is presented in Table 2, in which we indicate the telescope, its field of view (FOV), CCD matrix size, image resolution, filter and its central wavelength ($\lambda$), and full width at half maximum (FWHM).

The observing log is presented in Table 3, listing the begin-cycle time, the heliocentric ($r_h$) and geocentric ($\Delta$) distances, the phase angle of the object ($\alpha$), the position angle of the extended Sun-asteroid radius vector ($\theta_\odot$), the filter, the total exposure time during the night ($T_{exp}$), the number of cycles of observations obtained during the night ($N$), the mode of observations, and the telescope.

### 2.1 GTC 10.4-m telescope

We obtained CCD images of 2005 QN$_{173}$ on 2021 July 10, and low-resolution visible spectra on 2021 July 14, by using GTC telescope equipped with the Optical System for Imaging and Low-Resolution Integrated Spectroscopy (OSIRIS) camera-spectrograph (Cepa 2010). The OSIRIS detector is a mosaic of two Marconi CCDs operated in a 2 × 2 binned mode, with a readout speed of 200 kHz (gain of 0.95 e⁻/ADU; readout noise of 4.5 e⁻). Observational details are shown in Table 3.

On July 10, we obtained separate images using the Sloan $g'$, $r'$, $i'$ filters with an individual exposure time of 180 s. We did one $r'$, $g'$, $r'$, $i'$, and $r'$ sequence of images, taking three images each time using a small offset of 5 arcsec between them, with the telescope tracking at the comet's proper motion. Images were bias and flat-field corrected (using sky flats); all the images of the same filter were aligned and averaged combined to obtain the final images analysed in this paper. The object presents a conspicuous straight tail as shown in Section 3 but not a conspicuous coma (the FWHM of the asteroid in the optocentre is 1.3 arcsec while the field stars were 1.2 arcsec).

Three optical spectra of 2005 QN$_{173}$ were obtained on 2021 July 14 with the object at airmasses $1.2 < X < 1.3$. Each 900 s exposure time individual spectral image was obtained using the R300B grism and the 1.2 arcsec slit width oriented in parallactic angle. In this configuration, the resulting spectrum covers a wavelength range from 3600 to 7500 Å with a dispersion of 9.92 Å pix$^{-1}$ for a 1.2 arcsec slit. Spectral images were bias and flat-field corrected, using lamp flats. Since there is no conspicuous coma, and the slit is not oriented in the tail direction, only one-dimensional spectra were then extracted using an aperture of ±1.52 arcsec centred at the maximum of the intensity profile of the comet and collapsed to one dimension, wavelength calibrated (using Xe+Ne + HgAr lamps), and flux calibrated using the spectrophotometric standard star Ross 640. Two G-type stars from Landolt (1992) were also observed using the same configuration, one immediately before (SA112-1333 at $X = 1.2$) and another immediately after (SA115-271 at $X = 1.2$). They were used as solar analogues to obtain the normalized reflectance spectrum of the object, which is the average of the reflectance spectrum obtained with the extracted spectrum of each G-type star. As usual, the reflectance spectrum is obtained by dividing the spectrum of the object by the spectrum of the solar analogue star and normalized at 5500 Å.

---
[1] https://ssd.jpl.nasa.gov/

**Table 2.** Equipment used for observations.

| Telescope | FOV arcmin | CCD size px | Resolution, ″/px | Filter | $\lambda$ Å | FWHM Å |
|---|---|---|---|---|---|---|
| 10.4-m GTC | 7.8 × 7.8 arcmin² | 2048 × 4096 | 0.127 | g' | 4750 | 1530 |
|  |  |  |  | r' | 6300 | 1760 |
|  |  |  |  | i' | 7820 | 1510 |
| 6-m BTA | 6.1 × 6.1 arcmin² | 4600 × 2048 | 0.20 | g-sdss | 4650 | 1300 |
|  |  |  |  | r-sdss | 6200 | 1200 |
|  |  |  |  | R | 6420 | 790 |
| 2.5-m CMO | 10 × 10 arcmin² | 4296 × 4102 | 0.155 | B | 4347 | 1028 |
|  |  |  |  | V | 5379 | 888 |
|  |  |  |  | R | 6497 | 1256 |

**Table 3.** Observational circumstances of the data presented in this work. Information includes date, airmass ($X$), heliocentric ($r_h$) and geocentric ($\Delta$) distances, phase angle ($\alpha$), position angle of the projected anti-Solar direction ($\theta_\odot$), and the position angle of the projected negative heliocentric velocity vector ($\theta_{-V}$). Orbital values have been taken from JPL's HORIZONS system.

| Date, UT | $X$ | $r_h$, au | $\Delta$, au | $\alpha$, deg | $\theta_\odot$, deg | $\theta_{-V}$, deg | Filter/greed | $T_{\rm exp}$, sec | N | Mode | Telescope |
|---|---|---|---|---|---|---|---|---|---|---|---|
| 2021-07-10.15 | 1.446 | 2.393 | 1.897 | 24.05 | 246.56 | 246.65 | r' | 180 | 3 | image | GTC |
| 2021-07-10.15 | 1.399 | 2.393 | 1.897 | 24.05 | 246.56 | 246.65 | g' | 180 | 3 | image | GTC |
| 2021-07-10.16 | 1.357 | 2.393 | 1.897 | 24.05 | 246.56 | 246.95 | r' | 180 | 3 | image | GTC |
| 2021-07-10.17 | 1.321 | 2.393 | 1.897 | 24.05 | 246.56 | 246.95 | i' | 180 | 3 | image | GTC |
| 2021-07-10.17 | 1.288 | 2.393 | 1.896 | 24.05 | 246.56 | 246.95 | r' | 180 | 3 | image | GTC |
| 2021-07-14.15 | 1.281 | 2.395 | 1.855 | 23.50 | 246.56 | 246.91 | R300B | 2700 | 3 | spectra | GTC |
| 2021-07-15.98 | 1.380 | 2.397 | 1.835 | 23.30 | 246.57 | 246.66 | g-sdss | 280 | 7 | image | BTA |
| 2021-07-15.98 | 1.380 | 2.397 | 1.835 | 23.30 | 246.57 | 246.66 | r-sdss | 360 | 12 | image | BTA |
| 2021-07-15.99 | 1.392 | 2.397 | 1.834 | 23.30 | 246.57 | 246.66 | VPHG1200@540 | 1800 | 3 | spectra | BTA |
| 2021-07-17.96 | 1.364 | 2.398 | 1.824 | 23.16 | 246.58 | 246.66 | r-sdss | 2400 | 8 | imaPol | BTA |
| 2021-07-18.99 | 1.586 | 2.399 | 1.805 | 22.87 | 246.58 | 246.67 | V | 1000 | 5 | image | CMO |
| 2021-07-19.00 | 1.586 | 2.399 | 1.805 | 22.87 | 246.58 | 246.67 | R | 1000 | 5 | image | CMO |
| 2021-07-19.00 | 1.481 | 2.399 | 1.805 | 22.86 | 246.58 | 246.67 | B | 1000 | 5 | image | CMO |
| 2021-07-21.96 | 1.636 | 2.401 | 1.776 | 22.36 | 246.59 | 246.67 | V | 1000 | 5 | image | CMO |
| 2021-07-21.97 | 1.536 | 2.401 | 1.776 | 22.36 | 246.59 | 246.67 | R | 1000 | 5 | image | CMO |
| 2021-07-21.98 | 1.536 | 2.401 | 1.776 | 22.36 | 246.59 | 246.67 | B | 1000 | 5 | image | CMO |
| 2021-08-27.95 | 1.356 | 2.438 | 1.500 | 11.25 | 246.58 | 246.65 | V | 2000 | 4 | image | CMO |
| 2021-08-27.96 | 1.376 | 2.438 | 1.500 | 11.24 | 246.58 | 246.65 | B | 2000 | 4 | image | CMO |
| 2021-08-27.98 | 1.376 | 2.438 | 1.500 | 11.24 | 246.58 | 246.65 | R | 2000 | 4 | image | CMO |
| 2021-09-05.87 | 1.485 | 2.449 | 1.470 | 7.33 | 246.58 | 246.63 | R | 1200 | 4 | image | CMO |
| 2021-09-05.87 | 1.438 | 2.449 | 1.470 | 7.32 | 246.58 | 246.63 | V | 1200 | 4 | image | CMO |
| 2021-09-05.88 | 1.438 | 2.449 | 1.470 | 7.32 | 246.58 | 246.63 | B | 1200 | 4 | image | CMO |
| 2021-09-06.01 | 1.582 | 2.449 | 1.470 | 7.26 | 246.57 | 246.63 | R | 1200 | 4 | image | CMO |
| 2021-09-06.02 | 1.648 | 2.449 | 1.470 | 7.25 | 246.57 | 246.63 | V | 1200 | 4 | image | CMO |
| 2021-09-06.03 | 1.648 | 2.449 | 1.470 | 7.25 | 246.57 | 246.63 | B | 1200 | 4 | image | CMO |
| 2021-10-07.96 | 2.136 | 2.494 | 1.530 | 7.70 | 66.59 | 246.71 | R | 1200 | 4 | image | CMO |
| 2021-10-07.96 | 2.136 | 2.494 | 1.530 | 7.70 | 66.59 | 246.71 | B | 1200 | 4 | image | CMO |
| 2021-10-07.97 | 2.136 | 2.494 | 1.530 | 7.70 | 66.59 | 246.71 | V | 1200 | 4 | image | CMO |
| 2021-10-08.84 | 1.429 | 2.495 | 1.535 | 8.07 | 66.60 | 246.71 | V | 1200 | 3 | image | CMO |
| 2021-10-08.83 | 1.429 | 2.495 | 1.535 | 8.07 | 66.60 | 246.71 | R | 1200 | 4 | image | CMO |
| 2021-10-08.83 | 1.429 | 2.495 | 1.535 | 8.07 | 66.60 | 246.71 | B | 1200 | 4 | image | CMO |
| 2021-10-08.80 | 1.840 | 2.495 | 1.536 | 8.11 | 66.60 | 246.71 | g-sdss | 420 | 7 | image | BTA |
| 2021-10-08.81 | 1.840 | 2.495 | 1.536 | 8.11 | 66.60 | 246.71 | r-sdss | 780 | 13 | image | BTA |
| 2021-10-08.82 | 1.991 | 2.495 | 1.536 | 8.12 | 66.60 | 246.71 | VPHG1200@540 | 2400 | 4 | spectra | BTA |
| 2021-10-08.86 | 2.616 | 2.495 | 1.536 | 8.16 | 66.60 | 246.71 | R | 900 | 15 | imaPol | BTA |
| 2021-12-14.68 | 1.428 | 2.614 | 2.326 | 22.01 | 66.57 | 246.64 | V | 1000 | 5 | image | CMO |
| 2021-12-14.68 | 1.396 | 2.613 | 2.326 | 22.00 | 66.57 | 246.64 | R | 1000 | 5 | image | CMO |
| 2021-12-26.63 | 1.416 | 2.638 | 2.508 | 21.85 | 66.57 | 246.64 | R | 2700 | 9 | image | CMO |
| 2021-12-26.67 | 1.416 | 2.638 | 2.508 | 21.85 | 66.57 | 246.64 | B | 3500 | 5 | image | CMO |
| 2021-12-26.68 | 1.416 | 2.638 | 2.508 | 21.85 | 66.57 | 246.64 | V | 3500 | 5 | image | CMO |
| 2022-01-28.69 | 1.789 | 2.707 | 2.998 | 19.01 | 67.04 | 247.08 | R | 2000 | 4 | image | CMO |

## 2.2 BTA 6-m telescope

Quasi-simultaneous observations of the asteroid 2005 QN$_{173}$ were performed on 2021 July 15 and 17 and October 8 with Spectral Camera with Optical Reducer for Photometrical and Interferometrical Observations, a focal reducer decreasing the focal ratio at the prime focus of the 6-m telescope from (*f*/4) to (*f*/2.6) (see detail Afanasiev & Moiseev 2011; Afanasiev & Amirkhanyan 2012). Photometry and polarimetry of the comet were performed through broadband filters (see Table 3). To increase the signal/noise ratio, we applied binning of 2 × 2 to the photometric and polarimetric images and 1 × 2 to the spectroscopic frames. For spectral observations, we used the VPHG1200@540 grism and the long slit with dimensions of 6.1 × 1.0 arcmin$^2$. These provide an effective wavelength region of $\lambda = 3600 - 7250$ Å($\lambda = 4000 - 7000$ Åwithout a strong noise) and a dispersion of 0.91 Å px$^{-1}$. The spectral resolution is about 10 Å. Image scale is 0.2 × 0.4 arcsrc px$^{-1}$ for the spectral observations. We observed the spectrophotometric standard star BD + 25 4655 (Oke 1990) for absolute calibration. The telescope was tracked on the asteroid during the exposition. The spectral atmospheric transparency at the SAO was provided by Kartasheva & Chunakova (1978). The twilight morning sky was used for flat-field corrections of the photometric and polarimetric images. We applied the reduction procedure – bias subtraction, flat-field correction, and cleaning cosmic ray tracks in the standard manner, using IDL routines (e.g. Ivanova et al. 2019, 2021). The residual sky background was estimated with the use of an annular aperture. To perform an absolute flux calibration of the images, field stars were used.

To perform wavelength calibration, the spectrum of a He-Ne-Ar lamp was used. To provide flat-field corrections for the spectral data, we used a smoothed spectrum of an incandescent lamp. The observation nights were photometric with seeing near 1.5 arcsec. We used a high-resolution solar spectrum (Neckel & Labs 1984) to search for these possible emissions if they were present in the spectrum of the asteroid. The solar spectrum was transformed to the resolution of the spectrum by convolving with an instrumental profile and normalized to the flux from the asteroid. The procedure of continuum calculation was described in detail in our previous works (e.g. Ivanova et al. 2018, 2019, 2021, 2023). To estimate the instrumental polarization and the correction for the zero point of the position angle of the polarization plane, we observed the polarized and non-polarized standard stars with well-known large interstellar polarization taken from the lists of Hsu & Breger (1982), Schmidt, Elston & Lupie (1992), and Heiles (2000). The instrumental polarization was less than 0.1 per cent.

## 2.3 CMO 2.5-m telescope

Photometric observations of the asteroid 2005 QN$_{173}$ in *B, V*, and *R* filters were also carried out from July 2021 to January 2022. Absolute flux calibration of the asteroid images was carried out by measuring field stars. For this, we used catalogs the SDSS Photometric Catalogue, Release 12 (Alam et al. 2015) and ATLAS Refcat2 (Tonry et al. 2018b). The observations were carried out with a good seeing ∼1.6 arcsec. For all the photometric data, the standard data reduction was made. A detailed description of the processing

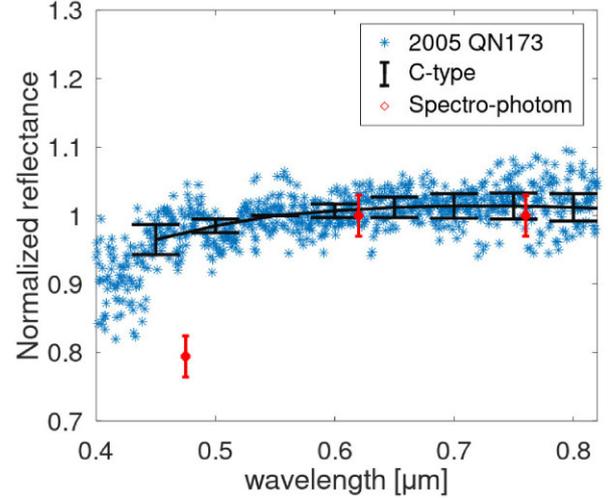

**Figure 1.** Reflectance spectrum of (248370) 2005 QN$_{173}$ obtained with GTC on July 14 compared with the typical spectrum of a C-type asteroid showing that it closely matched it. Overplotted (red dots) are the reflectances derived from the GTC colour photometry in $g'$, $r'$, $i'$ obtained on July 10. Notice the difference in the near-UV region (see text).

of all observational data can be found in Kornilov et al. (2016) and Ivanova et al. (2020).

## 3 RESULTS AND DISCUSSION

### 3.1 Spectroscopy

The reflectance spectrum obtained with GTC is analysed to determine the taxonomic class using the tool in M4AST[2] interface (Popescu, Birlan & Nedelcu 2012). The classification was performed in the framework of the Bus-DeMeo taxonomy (DeMeo et al. 2009). The result of the fitting clearly shows that the spectrum closely matched that of a C-type asteroid (see Fig. 1)

In this figure, we also plot the reflectances computed using the colours determined using GTC images on July 10 (($g' - r'$) = 0.76 ± 0.02; ($r' - i'$) = 0.12 ± 0.02). Assuming the colours of the Sun provided by Holmberg, Flynn & Portinari (2006) for the SDSS filters ($g' - r'$)$^{Sun}$ = 0.45 ± 0.02 mag, ($r' - i'$)$^{Sun}$ = 0.12 ± 0.01 mag, reflectances are computed in Popescu et al. (2018)

$$R_{aster}^{f1}/R_{aster}^{f2} = 10^{-0.4(C_{f1-f2}-C_{f1-f2}^{Sun})} \quad (1)$$

where *f*1 and *f*2 are two different filters, the corresponding colour is $C_{f1-f2}$, $C_{f1-f2}^{Sun}$ represent the colour of the Sun, $R_{aster}^{f1}$ and $R_{aster}^{f2}$ are the asteroid reflectances.

Notice that there is a significantly large difference in the reflectance computed using aperture photometry with respect to that obtained from the spectrum in particular in the region of the g' filter. It should be considered that the reflectance spectra of asteroids in the near-UV region are very sensitive to the used solar analogue star, as shown in Tatsumi et al. (2022), and we used G-stars that are not necessary solar analogues in this wavelength region. This can affect the shape of the reflectance spectrum below 0.5 $\mu$m, in particular, this effect can be strong below 0.45 $\mu$m as the star spectrum can be very different from that of the Sun. On the other hand, differences in the seeing between images can introduce some false magnitude variations in the measured brightness of the cometary coma (Licandro et al. 2000), and this introduces another uncertainty source. In this case, the difference between spectroscopic and photometric reflectance in the g-band is high (∼20 %), higher than those reported in Tatsumi et al. (2022), so a real colour variation between July 10 and 14 cannot be discarded.

---
[2]http://m4ast.imcce.fr/

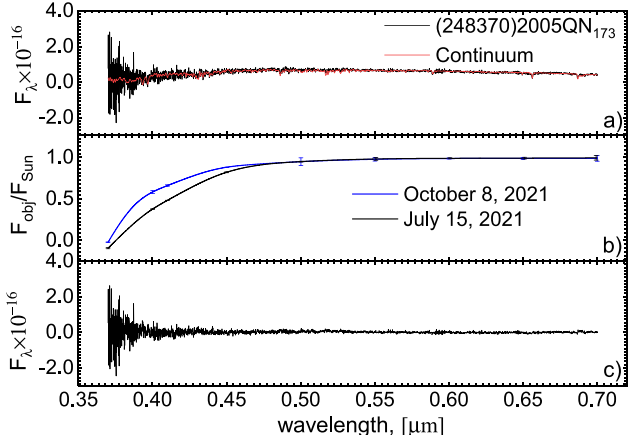

**Figure 2.** The spectrum of asteroid (248370) 2005 QN$_{173}$ in units of erg s$^{-1}$cm$^{-2}$Å$^{-1}$ derived on 2021 July 15: (a) the energy distribution in the observed spectrum (black line) and the transformed solar spectrum (red line); (b) the normalized spectral dependence of the dust reflectivity (black line—July 15 and blue line—October 8); (c) the spectrum of asteroid (248370) 2005 QN$_{173}$ without continuum component.

The change in the reflectance, along with the dispersion, shows the difference in the colour of the coma grains compared to the colour of the incoming solar radiation. Panel b in Fig. 2 presents the reddening of the asteroid continuum obtained at BTA in July. Usually, the slope (A'Hearn et al. 1984) or spectral gradient of reflectance $S'$ is used to quantify reddening, and its average value is expressed over a certain wavelength interval per 0.1 $\mu$m (Luu & Jewitt 1990). Reddening or bluer of dust (i.e. the colour excess of the scattered light compared to the incoming solar radiation) is indicated by positive or negative values, respectively. In order to define $S'$, we used expression:

$$S'(\lambda_1, \lambda_2) = \frac{20}{(\lambda_2 - \lambda_1)} \frac{(S_2 - S_1)}{(S_2 + S_1)}, \quad (2)$$

where $S_1$, $S_2$ are reflectivity for wavelength $\lambda_1$ and $\lambda_2$, respectively. In Fig. 2c, we do not see the presence of any gaseous emissions. This, in its turn, allows us to select areas on the reddening curve that can be represented by a polynomial of the first degree to determine the spectral gradient of reflectance $S'$. As we can see in Fig. 2b, starting from approximately 0.48–0.7 $\mu$m, the reddening curves for both dates coincide within the error limits and are linear. For these areas, we obtained the values of the one as about 3 per cent ± 0.2 per cent per 0.1 $\mu$m. Here, to obtain the energy distribution in the spectrum of the asteroid, we used a square diaphragm with a half-width of about 5000 km from the optocentre along the slit.

Although Fig. 2c shows the absence of any gaseous emissions, gassing is considered one of the possible mechanisms of asteroid activity (Jewitt, Hsieh & Agarwal 2015). Therefore, using the Haser model (Haser 1957), we calculated upper limits on the rate of production of the main cometary molecules, assuming that the useful signal of gas emissions does not exceed the noise. Also, we adopted a Gaussian function having an FWHM equal to the spectral resolution as an equivalent of the minimal measurable signal. So, the amplitude of the Gaussian was equal to the RMS noise level (Fig. 2c) for the observed fluxes through narrowband filters CN, C$_3$, C$_2$, CO$^+$, and NH$_2$ (Farnham, Schleicher & A'Hearn 2000). Because we used spectral observation, the area of the rectangular slit with a half width of 4 arcsec (about 5200 km) from the optocentre along the spatial coordinate was transformed into a circular aperture, and therefore, the observed flux is corrected. We use the values of the fluorescence efficiency (the g-factor), the parent and daughter scale lengths ($l_p$ and $l_d$), the power-low index $n$ dependent on the heliocentric distance ($r^{-n}$) from table 10 of Langland-Shula & Smith (2011). To obtain the actual L/N (luminosity per molecule) for the given heliocentric distance, the g-factor for CN was adopted from Schleicher (2010). The results are presented in Table 4.

**Table 4.** Upper limits for the main emissions in the spectrum of asteroid (248370) 2005 QN$_{173}$ derived on 2021 July 15.

| Emission band | Flux F × 10$^{-15}$, erg · s$^{-1}$ × cm$^{-2}$ | Production rate log(Q), mol s$^{-1}$ |
|---|---|---|
| CN(0-0) | <5.1 | < 23.9 |
| C$_2$($\Delta v = 0$) | <4.8 | < 23.9 |
| C$_3$($\lambda$4050 Å) | <2.1 | < 21.7 |
| NH$_2$ ($\lambda$6630 Å) | <0.2 | < 23.1 |
| CO$^+$ ($\lambda$4266 Å) | <2.0 | – |

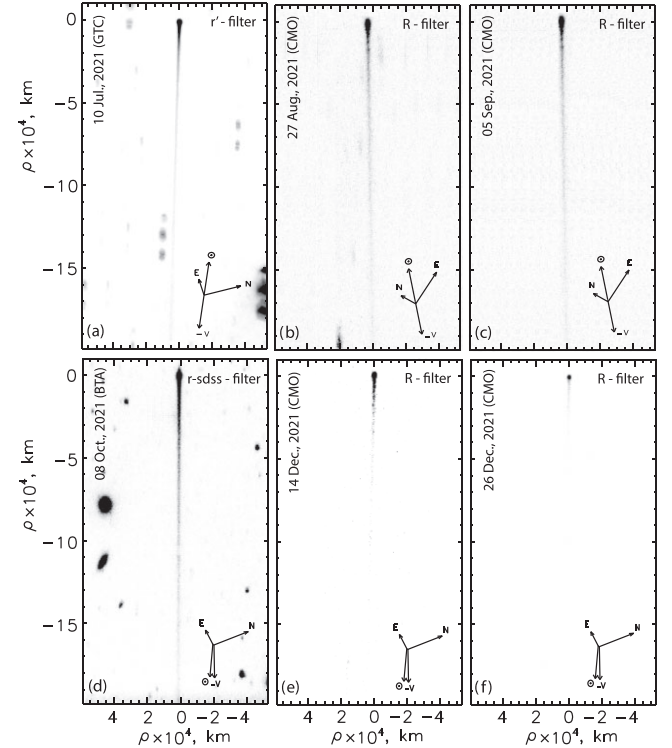

**Figure 3.** Intensity maps of (248 370) 2005 QN$_{173}$. The precise date and filter used are indicated on the corresponding panel. The directions to the North, the East, the Sun, and the negative heliocentric velocity vector of the comet as seen in the observer's plane of the sky are noted for each date.

### 3.2 Photometry

To analyse the morphological structure of the asteroid coma, we built intensity maps using data from photometrical nights. The results are shown in Fig. 3.

**Table 5.** The results of the photometry: aperture radius ($\rho$), apparent magnitude in $R$ filter ($m_R$), absolute magnitude in $R$ filter ($H_R$), colours $B-V$ and $V-R$, and parameter $A(0)f\rho$ in $B$, $V$, and $R$ filters.

| Date 2021 | $\rho$, km | $m_R$ | $H_R$ | $B-V$, mag | $V-R$, mag | $Af\rho(B)$, cm | $Af\rho(V)$, cm | $Af\rho(R)$, cm |
|---|---|---|---|---|---|---|---|---|
| July 10 | 5129 | $19.08 \pm 0.02$ | 14.67 | $0.98 \pm 0.04$ | $0.60 \pm 0.04$ | $15 \pm 1$ | $20 \pm 2$ | $25 \pm 1$ |
| July 15 | 5230 | $18.86 \pm 0.02$ | 14.54 | $0.67 \pm 0.04$ | $0.40 \pm 0.04$ | $26 \pm 1$ | $27 \pm 1$ | $27 \pm 1$ |
| July 18 | 5007 | $18.86 \pm 0.02$ | 14.59 | $0.65 \pm 0.03$ | $0.38 \pm 0.03$ | $27 \pm 1$ | $28 \pm 1$ | $28 \pm 1$ |
| July 21 | 4927 | $18.69 \pm 0.03$ | 14.47 | $0.69 \pm 0.04$ | $0.39 \pm 0.03$ | $30 \pm 2$ | $31 \pm 1$ | $31 \pm 1$ |
| August 27 | 4977 | $18.35 \pm 0.04$ | 14.84 | $0.62 \pm 0.03$ | $0.37 \pm 0.05$ | $24 \pm 1$ | $23 \pm 1$ | $23 \pm 1$ |
| September 05 | 4878 | $18.12 \pm 0.04$ | 14.81 | $0.69 \pm 0.03$ | $0.42 \pm 0.04$ | $23 \pm 1$ | $24 \pm 1$ | $25 \pm 2$ |
| September 07 | 4878 | $18.11 \pm 0.02$ | 14.75 | $0.64 \pm 0.03$ | $0.40 \pm 0.03$ | $24 \pm 1$ | $24 \pm 1$ | $24 \pm 1$ |
| October 07 | 4907 | $18.13 \pm 0.03$ | 14.67 | $0.70 \pm 0.04$ | $0.38 \pm 0.03$ | $26 \pm 2$ | $28 \pm 1$ | $27 \pm 2$ |
| October 08 | 4927 | $18.19 \pm 0.01$ | 14.71 | $0.70 \pm 0.03$ | $0.42 \pm 0.02$ | $25 \pm 2$ | $26 \pm 1$ | $27 \pm 1$ |
| October 08 | 4787 | $18.20 \pm 0.02$ | 14.71 | $0.60 \pm 0.04$ | $0.40 \pm 0.04$ | $27 \pm 1$ | $26 \pm 1$ | $26 \pm 1$ |
| December 14 | 4937 | $19.92 \pm 0.04$ | 14.93 | – | $0.33 \pm 0.08$ | – | $22 \pm 2$ | $21 \pm 1$ |
| December 26 | 5047 | $20.06 \pm 0.06$ | 14.90 | $0.94 \pm 0.08$ | $0.37 \pm 0.07$ | $16 \pm 2$ | $21 \pm 1$ | $21 \pm 2$ |

The asteroid coma is compact and bright. As one can see from Fig. 3, the asteroid demonstrates the tail structure during the whole observed period. Moreover, in October, a small elongation of the coma was detected in the antisolar direction. A similar morphology was observed in the objects in the previous activity period in 2016 (Chandler et al. 2021). Active asteroids (6478) Gault (Ivanova et al. 2019) and 133P/Elst–Pizarro (Jewitt 2012) demonstrated similar forms of compact coma and thin bright tail.

To estimate the dust productivity level, we calculated the $Af\rho$ parameter (A'Hearn et al. 1984). For this, first, the apparent magnitude of the asteroid was estimated within an aperture size equal to $\sim$5000 km on sky plane projection. Since we carried out observations within different filters, we had to have homogeneous results for the proper analysis. To convert the Sloan system magnitude to the Johnson-Cousins one, we used the transformation coefficients from Jester et al. (2005) for the $V$ filter and from Lupton et al. (2005) for the $R$ and $B$ filters. Using recalculated values of magnitudes, we computed dust productivity. The $Af\rho$ parameter is dependent on the phase angle of the object. Thus we applied the phase correction proposed by Schleicher, Millis & Birch (1998) for Comet Halley. The dust production measurement corrected for the phase angle is denoted as $A(0°)f\rho$. We should note that the nucleus was not subtracted in our dust productivity calculation. Also, using the measured apparent magnitudes of $QN_{173}$, we computed the magnitudes reduced to $r_h = \Delta = 1$ au and $\alpha = 0°$. For this, we applied the formula from Bowell et al. (1989) with the phase function coefficient $G = 0.15$. We collected calculated parameters based on photometrical observations obtained at all three telescopes in Table 5.

As one can see from Table 5, the asteroid primarily demonstrated red colour compared to the solar values taken from Willmer (2018) (0.63 and 0.39 for $B-V$ and $V-R$, respectively). Our computed apparent magnitudes, colours, and dust productivity are close to the ones obtained by Hsieh et al. (2021) and Novaković et al. (2022). We should notice that Hsieh et al. (2021) calculated the colours in the Sloan system. The nominal values and trends are close to those presented in Hsieh et al. (2021) and Novaković et al. (2022). Both measurements demonstrate an increase in brightness from July 10 to 18. In our observations, it continues to increase to July 21, while results from (Hsieh et al. 2021) show a similar value within the error bars. The general trend of calculated magnitudes based on all our observations has decreased character. Our measurements of colours are also comparable to ones obtained for other active asteroids (see, for instance, Jewitt et al. (2019); Borysenko et al. (2020); Carbognani & Buzzoni (2020); Ivanova et al. (2020); Lin et al. (2020); Carbognani, Buzzoni & Stirpe (2021); Devogèle et al. (2021), and references therein).

In general, the dust productivity of active asteroids (or main-belt comets) is significantly lower than that of comets, except for Jupiter family comets at small heliocentric distances. Studies of main-belt comet P/2012 T1 (PANSTARRS) provided by Hsieh et al. (2013) showed approximately the same level of dust productivity. The $Af\rho$ parameter calculated for P/2010 R2 (La Sagra) was slightly higher at similar heliocentric distances but still less than 100 cm (Hsieh et al. 2012). Photometric investigations of 238P/Read (Shi et al. 2019) revealed a low level of dust productivity (up to 10 cm in $Af\rho$ proxy). Borysenko et al. (2020) reported about even a lower value of the $Af\rho$ parameters for these two objects and for other main-belt comets, e.g. 311P/PANSTARRS and 331P/Gibbs.

### 3.3 Colour and polarization distribution

Using the photometric observations from the 6-m telescope of the asteroid and the technique described by Ivanova et al. (2019, 2021, 2023), we created the $g-r$ colour map (Fig. 4a,b) to analyse the dust properties in the coma and the detected tail. An error in the colour measurements varied from $0.02^m$ to $0.1^m$ with distance from the optocentre. From the colour map based on July observations, one can see that the near-nucleus area of the asteroid (up to $\sim$5000 km) has a red colour, on average, $\sim 0.5^m$. In general, the colour of the tail becomes bluer with increasing distance from the nucleus suggesting an evolution of the dust particles. We detected strong colour variations along the tail (up to $\sim$50 000 km) that are most probably associated with the moving of ejected material (or big fragments) and the rotation of the object. The different colours of big fractions can be a result of the composition of the ejected material that can differ from the average colour of the surface. In addition, the concentration and distribution of these clumps along the tail can vary, resulting in regions of different colours. Similar variations have also been detected in the polarization measurements, albeit to a lesser extent. Analysis of the October observations showed that we have an opposite situation, and the near-nucleus area of the asteroid has a bluer colour, on average, $\sim 0.4^m$, and the colour of the tail becomes redder with increasing distance from the optocentre. The variation of colour, along the tail, was detected in October too, but it was weaker.

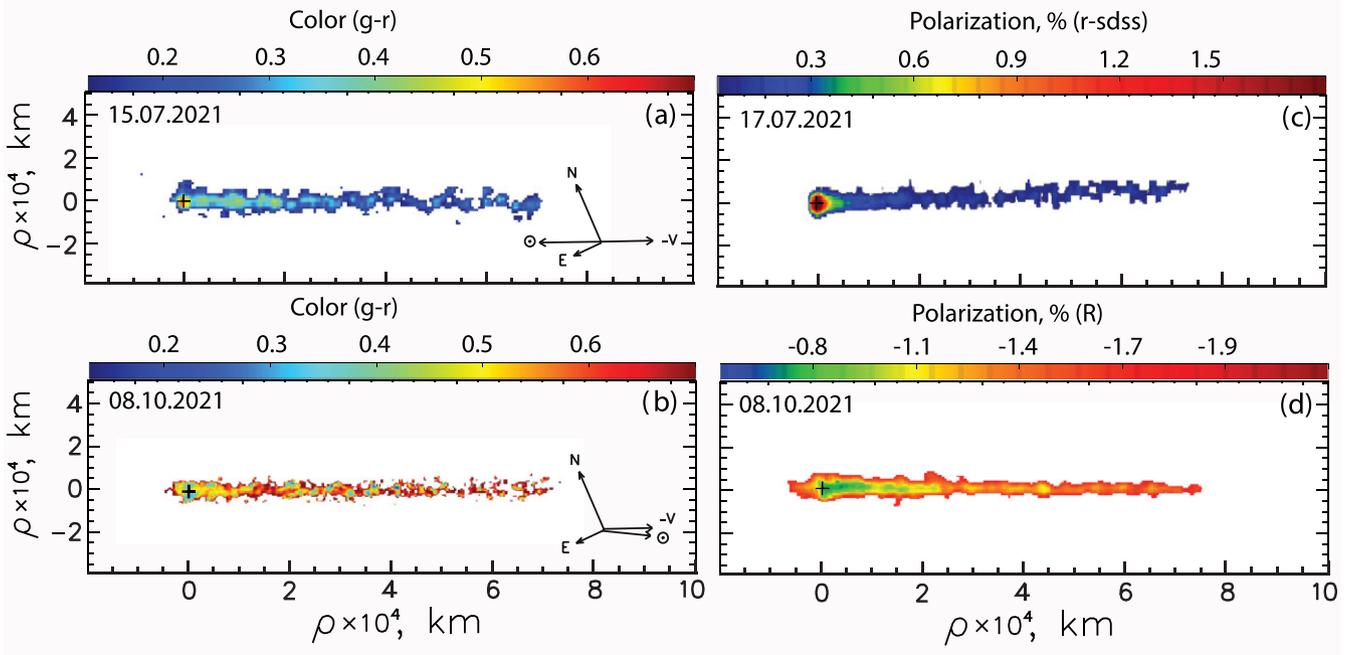

**Figure 4.** Distribution of colour $g-r$ (on the left) and linear polarization degree (on the right) over the coma and tail of asteroid (248370) 2005 QN$_{173}$ obtained in July (top panels) and October (bottom panels) at phase angles of $23.16°$ and $8.16°$, respectively. The directions to the North, the East, the Sun, and the negative heliocentric velocity vector of the comet as seen in the observer's plane of the sky are noted for each date.

For the first time, in an active asteroid, the linear polarization over the coma was observed, and detailed polarized maps (see Fig. 4c,d) in the *r-sdss* and *R* filters were built to analyse the dust properties in the coma and along the tail. The red colour shows areas with a high degree of polarization, while the blue colour presents a low polarization degree (in absolute values). For observations obtained in 2021 July in all areas near the optocentre of the asteroid, the polarization degree is positive, which means that the plane of polarization is perpendicular to the scattering plane. Fig. 4c shows that the polarization gradually decreases (in absolute value) from the optocentre to the tail. In the opposite situation, we can see from observations obtained in October (see Fig. 4d), the value of linear polarization increased with distance from the optocentre. The errors in the measured polarization degree of the comet varied between 0.15 per cent and 1.4 per cent.

A comparison of the colour and polarization maps in Fig. 4 shows that there is a relationship between the changes in colour and the degree of polarization for both dates of observations. In July, the near-nuclear region is characterized by a high degree of polarization ($\sim$1.2 per cent) and red colour near ($0.5^m$). The colour in the tail region varies from $0.4^m$ to $0.15^m$ and represented low linear polarization (0.2–0.02 per cent). In October, the colour and polarization near the optocentre are lower near $0.4^m$ and $-0.2$ per cent, and the tail is characterized by a high degree of polarization $-(0.9-1.5$ per cent) and colour changes from $0.35^m$ to $1.5^m$.

Such a relationship between colour and polarization indicates dust particles of various compositions and sizes released from the surface of the active asteroids. To investigate whether there are any polarimetric and colour trends near the optocentre and along the tail, we have taken scans from the photometric centre (see Fig. 5). We measured the colour and polarization within the coma area of $2 \times 2$ px$^2$, starting from the photometric centre of the asteroid.

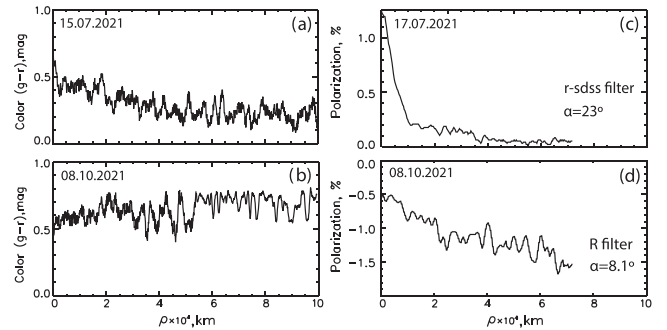

**Figure 5.** The observed radial profiles of $g-r$ colour (panels a and b) and polarization (panels c,d) across the coma of asteroid (248370) 2005 QN$_{173}$ as a function of the distance from the nucleus. The top panels correspond to July observations and the bottom ones to the October set.

Fig. 6 shows the mean degree of polarization of the asteroid and the phase dependence of polarization of different C-class main-belt asteroids. It is evident that the degree of polarization of QN$_{173}$ is typical of C-type asteroids.

### 3.4 Numerical modelling of dust particles' characteristics

We modeled the light-scattering properties of dust particles by assuming that they are solid with the effective refractive index of $m = 1.7 + i0.005$. Such a refractive index was derived for meteorite powder representing a C-type asteroid (Roush 2003). Particle shapes were generated using the Voronoi partitioning method that creates irregularly shaped particles (Markkanen et al. 2015). The particle model parameters were selected such that the particles are solid with the roughness length scale $\sim 0.075 \, r$, where $r$ is the mean radius of the particles.

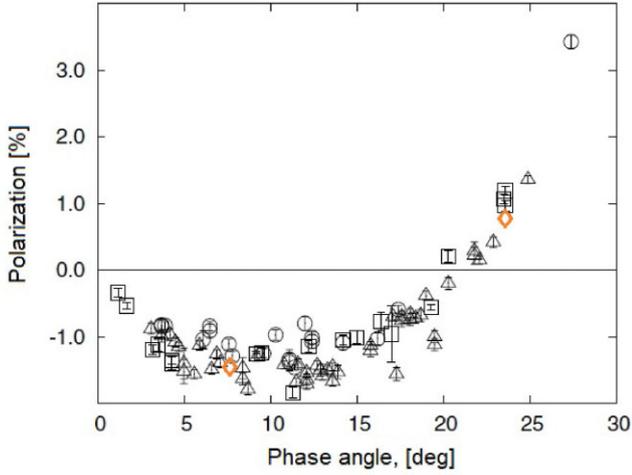

**Figure 6.** Phase dependence of the linear polarization degree for different C-class main-belt asteroids. The observations indicated by circles, triangles, and squares are measurements of Cb-, Ch-, C-type objects (Gil-Hutton & Cañada-Assandri 2012) and diamonds (our observations of (248370) 2005 QN$_{173}$), respectively.

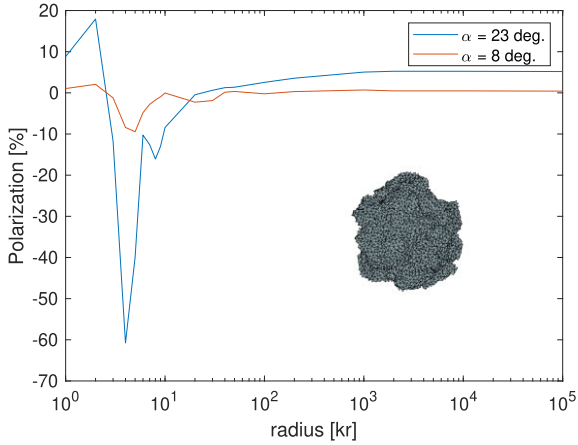

**Figure 7.** Modelled degree of linear polarization at two phase angles as a function of particle size. An example of the discretized particle shape model is also shown.

Two different numerical methods were used to solve the light-scattering properties of the model particles. The particle size is characterized by the size parameter $kr$, where $k$ is the wavenumber. For particle sizes $kr \leq 30$, we used the surface-integral-equation (SIE) method for the PMCHWT formulation. More details on the SIE method and its implementation can be found in Ylä-Oijala et al. (2014) and the references therein. Since the SIE method solves the Maxwell equations, it cannot be applied to particles much larger than the wavelength due to computational constraints. Therefore, for particles larger than $kr > 30$, we applied SIRIS4 code based on the geometric optics approximation using inhomogeneous plane waves (Lindqvist et al. 2018).

Fig. 7 shows the modelled degree of linear polarization of unpolarized light as a function of particle size parameters $kr$. For the SIE computations up to $kr = 30$, we generated 10 random shapes for each size and averaged each shape and size over 128 random orientations and 361 scattering planes. For SIRIS4 computations, 10 shapes were averaged using 10M rays for each shape.

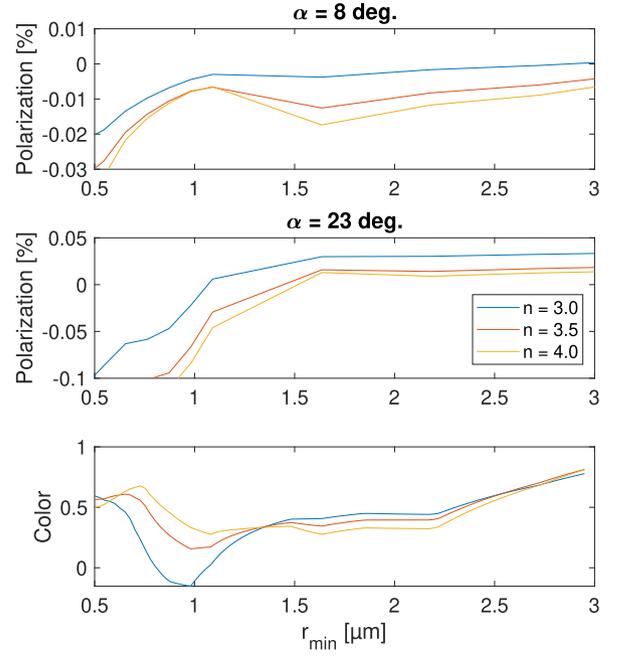

**Figure 8.** Modeled degree of linear polarization at two phase angles and colour for different power-law size distributions with index $n$.

As seen from Fig. 7, the value of polarization increases with increasing particle size if the particles are large enough. Such behaviour is consistent with the observed polarization maps in Fig. 4 as large particles are expected to populate the near nucleus regime due to the effect of radiation pressure. Such a dynamical sorting effect can be easily conformed by drawing a grid of synchrones and syndynes using a simple Finson & Probstein model (Finson & Probstein 1968).

For more detailed analysis, we plot in Fig. 8 the polarization degree and colour averaged over the power-law size distributions defined by the power-law index $n$ and the minimum $r_{min}$. The maximum radius was constant $r_{max} = 1$ cm. The refractive index was assumed to be the same for both wavelengths 465 and 685 nm in the colour computations. The modeled colour was indistinguishable for both phase angles.

Comparing the modelling results in Fig. 8 and the observations in Fig. 4, it is evident that large particles are concentrated around the nucleus whereas smaller ones dominate in the tail. The correlation between polarization and colour indicates the absence of smaller than micron-sized particles whereas anticorrelation would indicate the presence of small particles. Polarization increases with decreasing power-law index $n$ for all modeled minimum particle sizes $r_{min}$. The dependence of colour with $n$, however, depends on the minimum particle size $r_{min}$. For $r_{min} < 1.35\,\mu m$ colour increases with $n$ resulting in anti-correlation between the colour and polarization, whereas they are correlated for $r_{min} > 1.35\,\mu m$. We note, however, that it is not possible to derive the exact particle size distribution from the data as the polarization and colour also depend on the refractive index and the scale of surface roughness, which are unknown properties of dust in active asteroids. Changing the refractive index slightly would not change the overall behaviour of the polarization with respect to size. For compact moderately absorbing particles, the negative polarization branch is created by the micron-size particles and the polarization increases with size. Hence, changing the refractive index would slightly affect the retrieved $r_{min}$ value. Increasing the imaginary part of the refractive index would eventually remove the negative branch of polarization, whereas decreasing it would increase the geometric albedo of the asteroid surface too much.

### 3.5 Monte Carlo modelling of the dust tails

In order to retrieve the duration of the activity, the ejected mass, and the dust physical parameters from the observed images, we used the Monte Carlo code described, e.g. in Moreno (2022). Only the r-Sloan images were used, as they are possibly the less contaminated from gaseous emissions (although no gas emissions were detected).

Owing to the small size of the asteroid, its gravity can be neglected in our dynamical computations. In these circumstances, the trajectories of the particles (assumed spherical) are Keplerian. Their dynamics are driven by the gravity of the Sun and the solar radiation pressure only through the so-called $\beta$ parameter, defined as $\beta = C_{pr}Q_{pr}/(2\rho r)$, where $C_{pr} = 1.19 \times 10^{-3}$ kg m$^{-2}$, and $\rho$ is the particle density. Following Hsieh et al. (2021), we use $\rho = 1400$ kg m$^{-3}$, consistent with a C-type asteroid (Britt et al. 2002). Although for small asteroids, e.g, Rhygu and Bennu, the density is slightly smaller (~1190 kg m$^{-3}$, see Watanabe et al. 2019; Rozitis et al. 2020), we have preferred to keep the above value for the purposes of comparison with previous work. For particle radii $r > 0.25\,\mu$m, the radiation pressure coefficient is $Q_{pr} \sim 1$ (Burns, Lamy & Soter 1979). In the Monte Carlo process, each particle in the distribution is assumed to be ejected isotropically with a certain velocity. Then, its trajectory is computed from the ejection time till the observing time, and its position is then projected to the sky or (N, M) plane. Its brightness is a function of its size and the geometric albedo, which we have taken as $p_v = 0.054$ as reported by Mainzer et al. (2019), the same value assumed by Hsieh et al. (2021). In the Monte Carlo procedure, the ejection of a large number ($\gtrsim 10^7$) of particles is simulated at each time-step from the start to the end of the activity. A linear phase correction of 0.03 mag deg$^{-3}$ is applied to the particle phase function. The final synthetic brightness image is computed as the contribution of all the sampled particles, being a function of the mass loss rate and the size distribution. In addition, the nucleus brightness contribution is also taken into account, being characterized by its radius, $R_n$, assumed at having the same geometric albedo and linear phase coefficient as the particles.

The size distribution of the particles is assumed to be very broad, having initially a minimum at $r = 1\,\mu$m and a maximum at $r = 1$ cm, being distributed as a power-law of a certain index $\kappa$. After some experimentation with the code, the minimum size was set, however, to $r_{min} = 5\,\mu$m (see below). The dust mass loss rate is assumed as being a Gaussian function having a maximum $(dM/dt)_0$, FWHM, and a time of maximum ejection of $t_0$. The particle ejection velocities are parametrized as usual as $v = v_0\,\beta^\gamma$, which, for simplicity, we assumed as time-independent. We are finally left with six adjustable parameters in our model, namely $(dM/dt)_0$, $t_0$, $v_0$, $\gamma$, $\kappa$, and $R_n$. All the other physical parameters are always fixed at the values given above.

To find the best fit for the observations, we used a minimization technique based on the downhill simplex method (Melder & Mead 1965). The quantity to be minimized is the standard deviation of the model brightness with respect to the observed brightness, the procedure being summed up among all the available images. The best-fitting parameters were found as $(dM/dt)_0 = 2.3$ kg s$^{-1}$, $t_0 =$ JD 2 459 327 (2021 April 22, 2021, i.e., 23 d before perihelion), FWHM = 205 d, $v_0 = 12$ cm s$^{-1}$, $\gamma = 0$, $\kappa = -3.4$, and $R_n = 1.3$ km. To show the quality of the fits, a comparison of the observed images (left-hand panels) and the corresponding best-fitting images (right-hand panels) is shown in Fig. 9. It is seen that the images are closely reproduced, and, in particular, the perspective antitail on October 10 is clearly shown in the simulations. The brightness levels across the thin linear structures are shown in Fig. 10, where the agreement between the observations and the model is very good.

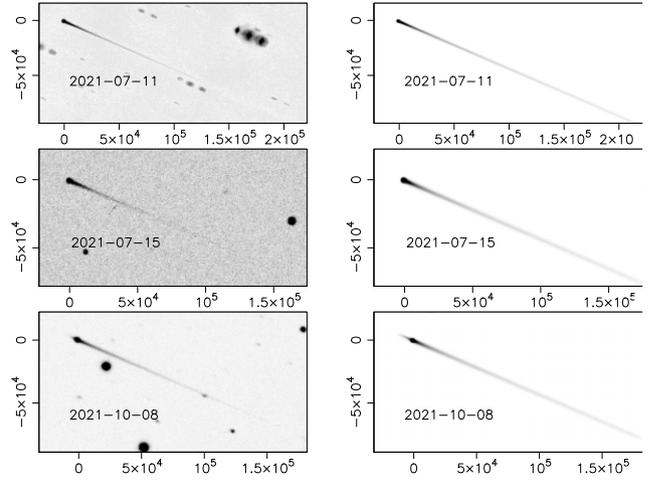

**Figure 9.** Comparison of the observed images (left-hand panels) and corresponding best-fitting synthetic images (right-hand panels), for the indicated dates. All graphs are labelled in kilometres projected on the sky plane relative to the nucleus position. North is up, and East is to the left in all images.

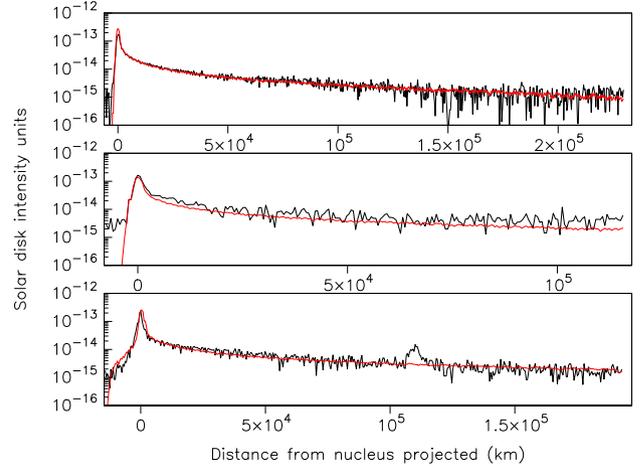

**Figure 10.** Brightness levels across the thin linear structures, as a function of the distance to the nucleus projected. Black lines are the observations, and red lines are the model. The upper, middle, and lower panels correspond to scans obtained from the images shown in Fig. 9 (2021 July 11, 2021 July 15, and 2021 October 08, respectively).

The long-lasting activity of this MBC (FWHM = 205 d, i.e. ~7 months) is not surprising, as it has been found recurrently active, at least during one previous apparition, from archive images (Chandler et al. 2021; Hsieh et al. 2021).

The perspective antitail observed on October 8, 2021, is populated by particles smaller than ~10 $\mu$m in size. To demonstrate this, Fig. 11 displays the synthetic images obtained using the best-fitting parameters except for the minimum radius, which is set to 1, 5, and 10 $\mu$m, in the upper, middle, and lower panels, respectively. For sizes as small as $r_{min} = 1\,\mu$m, the anti-tail would be too bright, and for sizes as large as $r_{min} = 10\,\mu$m the anti-tail would essentially vanish.

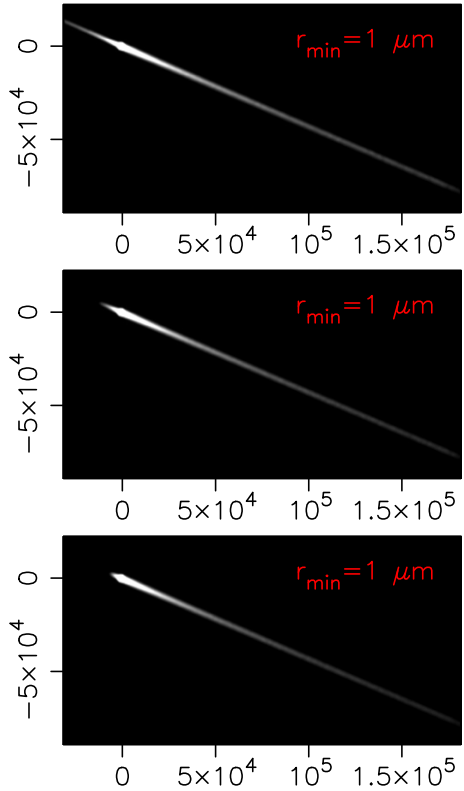

**Figure 11.** Synthetic images obtained using the best-fitting parameters except for the minimum radius, which is varied from 1 to 10 $\mu$m, as labelled. The $x$- and $y$-axis are labelled in kilometres projected on the sky relative to the nucleus position. North is up, and East to the left in all images. The upper, middle, and lower panels correspond to scans obtained from the images shown in Fig. 9 (2021 July 11, 2021 July 15, and 2021 October 08, respectively).

The minimum particle size was then finally set to 5 $\mu$m, which provides a reasonable fit of the images at all dates.

The total dust mass ejected until the latest observation of 2021 October 10 is $4.2 \times 10^7$ kg, with a maximum rate of 2.6 kg s$^{-1}$. These dust losses are similar to other studied MBC's. Thus, just to mention a few, Moreno et al. (2013) inferred $\sim 10^7$ kg for P/2012 T1 (Pan-STARRS), with a maximum loss rate of 1 kg s$^{-1}$ and 4 kg s$^{-1}$ for P/2010 R2 (La Sagra), with activity lasting at least 3 months (Moreno et al. 2011). Similarly, Jewitt et al. (2014) found $dM/dt = 0.2$–$1.4$ kg s$^{-1}$ for the long-lasting, recurrently active, MBC prototype, 133P/Elst-Pizarro. Another common feature is the slow ejection speeds, mostly in the range of a few tens of metres per second, as in this case. Hsieh et al. (2021) obtained dependence of the kind $v = 1.9/\sqrt{r}$ m s$^{-1}$, (with $r$ in microns), which results in speeds of $v = 20$ cm s$^{-1}$ at 100 $\mu$m or about 5 cm s$^{-1}$ for mm-sized particles, i.e., in line with ours, although our best fit indicates no dependence on size ($\gamma = 0$). On the other hand, the size distribution power index ($\kappa = -3.4$) is within the usual range found for these objects. Finally, the nucleus size ($R_n = 1.3$ km) is close to that found by Hsieh et al. (2021) of $R_n = 1.6 \pm 0.2$ km.

# 4 MODELLING OF THE NEAR FUTURE ORBITAL EVOLUTION IN THE MAIN-BELT

In this section, we focused on the near future evolution of the 2005 QN$_{173}$ orbit and the possible approach to the comets, which can be active in these heliocentric distances. It is a methodological approach based on the numerical integration of the object orbits and the comparison of the mutual distances of the orbital arcs of these objects, without using observational data. Novaković et al. (2022) performed a numerical integration of the 2005 QN$_{173}$ orbit for 100 Myr backward in time, and they investigated the long-term dynamical stability of this object. From this numerical integration (specifically from the duration of the numerical integration), it can conclude that this body was not gravitationally captured from a cometary orbit into the Main-belt.

The nominal osculating orbital elements of the asteroid 2005 QN$_{173}$ orbit, needed for numerical integration, were taken from the JPL Small-Body Data base Lookup.[3] To cover possible scenarios of the nominal orbit evolution of 2005 QN$_{173}$, we created 100 clones of its orbit, representing the statistical uncertainty. For this, we used the method developed by Chernitsov, Baturin & Tamarov [1998; for details, see Tomko & Neslušan (2019)]. The numerical integration of the orbits was performed by using the integrator RA15 (Everhart 1985) within the MERCURY package (Chambers 1999). The gravitational perturbations of the eight planets (from Mercury to Neptune) are considered in this integration.

The evolution of 2005 QN$_{173}$ orbit and its clones forward in time for 100 kyr are shown in Fig. 12. The solid red curve shows the evolution of the orbital elements of 2005 QN$_{173}$, and the green curves show the evolution of the clones. No differences between the nominal orbit and the orbits of the clones are visible. Small differences are seen only after 80 kyr from the beginning of integration. The changes of the orbital elements over time have the profile of a sinusoid curve with a period of maximum (minimum) changes of around 12 000 yr. The evolution of perihelion distance and eccentricity are in antiphase but with the same period of changes. The dynamical analysis into the past performed by Novaković et al. (2022) showed that 2005 QN$_{173}$ is an unstable object and is trapped inside an orbital mean-motion resonance (MMR), centred around 3.076 au, that it is 11:5 MMR with Jupiter. Our biggest planet has a great influence on the bodies in the Main-belt. The same applies to 2005 QN$_{173}$ and its trajectory into the future.

The minimum orbit intersection distance (MOID) between the orbital arcs of 2005 QN$_{173}$ and Jupiter during the whole investigated period are shown in Fig. 13. Profile of the evolution of MOID follows the same course as the evolution of the eccentricity for object 2005 QN$_{173}$ (Fig. 12b). The period of these approaches of 2005 QN$_{173}$ orbit to the Jupiter orbit is the same as the period of the changes of the orbital elements. Thus, Jupiter has a significant influence on the evolution of the orbit profile and its location in the Main-belt. Jupiter controls the change of the orbital parameters. We can also conclude that due to its influence, there is a very low dispersion of the clones of the nominal orbit.

## 4.1 Close approaches of comets

Comets or even asteroids can become active at these heliocentric distances, i.e. particles escape from their nuclei. Subsequently, they can form streams of meteoroids through which other bodies can pass. These particles can disrupt the surface any body and thereby potentially trigger its activity. But that requires a long series of observations and modelling. However, the distance between the orbits of the body of our interest and some comets is important. We calculated approaches of the 2005 QN$_{173}$ orbit and the orbits of the sample of the 464 short-periodic comets (labelled from 1P/Halley to 380P/PANSTARRS, including all 66 fragments of comet 73P/Schwassmann-Wachmann 3, 2 fragments of 141P/Machholz 2, 3 fragments of 205P/Giacobini, and 9 fragments of 332P/Ikeya-Murakami), and we tracked their mutual positions.

---

[3] https://ssd.jpl.nasa.gov/tools/sbdb_lookup.html#/?sstr = 248370

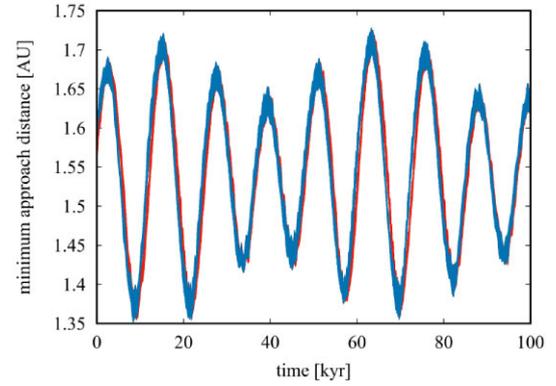

**Figure 13.** Evolution of the minimum distance between the orbital arcs of 2005 QN$_{173}$ and Jupiter during the next 100 kyr. The minimum distance of the post-perihelion arc is shown by the red curve and that of the pre-perihelion arc by cyan.

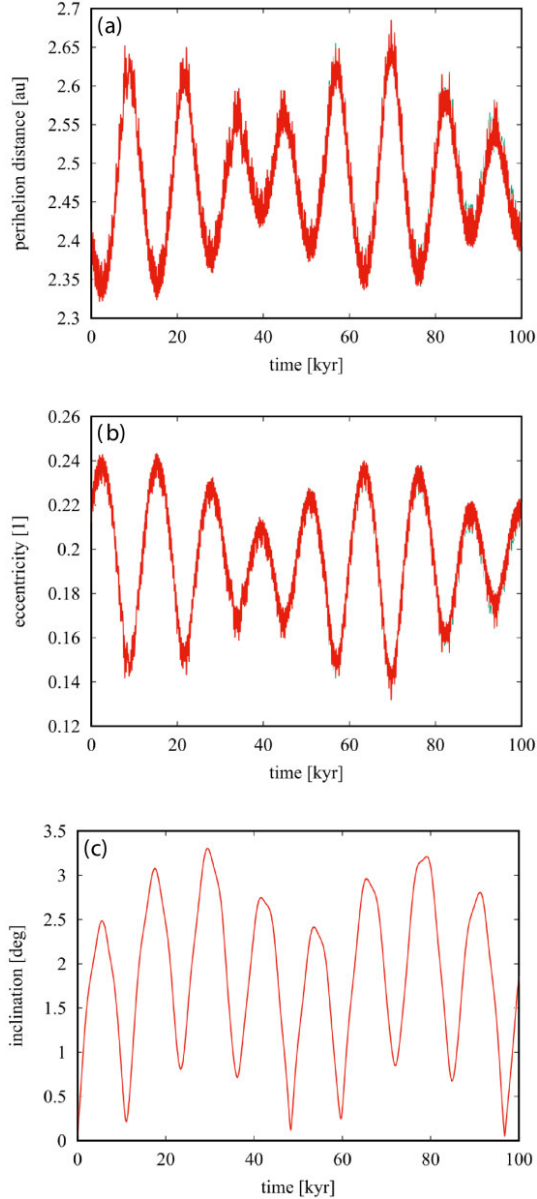

**Figure 12.** The evolution of the perihelion distance (a), the eccentricity (b), and the inclination to the ecliptic plane (c) of the orbit of 2005 QN$_{173}$ (red curve), and the orbits of the set of its clones (green curves) for up to 100 kyr into the future.

The first step in determining the future mutual distances between the orbits of the objects is a numerical integration of their orbits. In this case, we also made a numerical integration of the orbits 2005 QN$_{173}$ and above mentioned the sample of the comets 4 kyr into the future. Integrator RA15 was also used in this integration. When the integration into the future was done, we selected those comets that moved in orbits approaching 2005 QN$_{173}$'s orbit within 0.02 au (with MOID smaller than 0.02 au). We chose this value to guarantee that there will be a close approach or even a crossing of the orbits. Despite the fact that we have performed the integration for 4 kyr into the future, we focused on the first 100 yr since the beginning of the integration. The activity of 2005 QN$_{173}$ cannot be reliably predicted. It was observed for the first time on 2021 July 7 by Tonry et al. (2018a), and later, it was found on archival observational images on 2016 July 22 by Chandler et al. (2021).

We divided all approaches according to the orbital arc on which the approaching took place: on the post-perihelion arc, pre-perihelion arc, and on both arcs. Of the total number of 464 comets, the orbits of 8 comets approach the 2005 QN$_{173}$ orbit. Four comets only on the post-perihelion arc (73P-L, 73P-AL, 120P), one comet on the pre-perihelion arc (64P), and five comets on both arcs (198P, 253P, 278P, 298P). We took into account those comets whose MOID is at least once less than 0.02 during the studied period. Actually, MOID was much lower for all of these comets.

Two comets have observed and confirmed activity: 64P/Swift-Gehrels and 120P/Mueller 1. Kelley et al. (2019) discovered a −2.7 magnitude outburst of comet 64P occurring around 2018 August 14. Secular multiwavelength broadband photometry of the 64P was performed by Xu et al. (2022). Their aims were also to research the physical properties and activity of 64P, and they discovered mini-outbursts on 2019 January 3. Our dynamical modelling reveals that, in time, observed activity was MOID between 2005 QN$_{173}$ and comet 64P equal 0.01 au. The MOID between these two objects is relatively low, less than 0.01 au for a period of 62 yr from the beginning of the modelling.

On 2021 August 1, an apparent outburst of comet 120P/Mueller 1 was discovered in Zwicky Transient Facility photometry by Kelley (2021). The outburst was subsequently confirmed on August 9. The comet was at 2.55 au from the Sun and 3.00 au from the Earth. According to our dynamical modelling of the orbit evolution, MOID between comet 120P and asteroid 2005 QN$_{173}$ is even smaller than in the case of comet 64P. The MOID between these two objects is less than 0.005 au after 42 yr from the beginning of the orbit modelling.

Very low MOID and similarity of orbital elements may indicate a common 'genetic' origin. In addition to photometric and spectroscopic observations, numerical methods of the orbits' similarity are also helpful. However, the probability that two objects are actually genetically related or have a common progenitor is relatively low. Vokrouhlický et al. (2022) studied extremely young asteroid pair (458271) 2010 UM26 and 2010 RN221. These two asteroids have similar heliocentric orbital elements. They analysed the conditions of its origin and determined its age. Archived and new astrometric observations were used with a modelling of the orbital convergence. Using a large number of possible clone variants of (458271) 2010

UM26 and 2010 RN221, they found that they all converge in a narrow time interval around March 2003. They have extremely tight minimum distances ($\leq 1000$ km) and minimum relative velocities ($\leq 3$ cm s$^{-1}$). Quasi-satellite captures mean that the possible age uncertainty of this pair might extend to the 1960s. Even in such a case, future astronomical observations are needed.

The similarity of the orbits and, hence, their possible relationship can be evaluated by using the Southworth & Hawkins (1963) D$_{SH}$ discriminant. The D$_{SH}$ has been widely applied in identifications of meteor streams and in other studies on genetic affinity among meteoroids, asteroids, and comets (Sekanina 1970a, b, 1973, 1976; Lindblad 1971a, b, c; Jopek 1986). Lindblad & Southworth (1971) applied this function in the studies of asteroid families. In our case, the determining of the similarity between two orbits, 2005 QN$_{173}$ and 10 approaching comets, was negative. The values of D$_{SH}$ discriminant were very high, which does not indicate their closer connection. We can conclude that these objects are not genetically related, despite the very close distance of their orbits for a relatively long time. The possible approach and the similarity of the orbits we can consider as random.

## 5 CONCLUSIONS

We provided a complex analysis of observations of active asteroid (248370) 2005 QN$_{173}$ obtained from July 2021 to January 2022.

(i) We detected no gas emissions in the asteroid spectrum obtained in July and October; normalized reflectivity gradient was estimated as about 3 per cent $\pm$ 0.2 per cent per 0.1 $\mu$m; the spectrum of the asteroid closely matched that of a C-type asteroid;

(ii) Analysis of photometrical data revealed the presence of asteroidal activity from July 2021 to January 2022 as the compact coma and the long dust tail; observable activity was decreased to the end of December; results obtained in October demonstrated coma elongated in the opposite direction to the Sun;

(iii) Asteroid demonstrated redder colour compared to the Sun one in $B-V$, $V-R$ diapasons; the colour values are close to the C-type asteroid ones; calculated dust productivity, $Af\rho$ proxy, was about $\sim$26 cm in $R$ filter with decreasing in December; dramatic changes in dust productivity obtained in different filters were not detected;

(iv) We detected the spatial variations of the colour and polarization over the coma: the $g$–$r$ colour changes from about 0.2$^m$ to 0.7$^m$ in July and October based on data obtained at 6-m telescope; the linear polarization degree varies from about 1.2 per cent to 0.2 per cent and from $-0.2$ per cent to $-1.5$ per cent at the phase angle of 23.2° and 8.16°, respectively;

(v) Based on modelling of dust particles' characteristics using photometric and polarimetric results, it is shown that large particles are concentrated around the nucleus whereas smaller ones dominate in the tail. Further, the correlation between polarization and colour indicates an absence of submicron-sized particles, whereas anticorrelation would indicate the presence of submicron particles.

(vi) Best-fitting parameters to the observations based on the Monte Carlo modelling of the dust tail were computed $(dM/dt)_0 = 2.3$ kg s$^{-1}$, $t_0$ = JD 2 459 327 (2021 April 22, i.e. 23 d before perihelion), FWHM = 205 d, $v_0 = 12$ cm s$^{-1}$, $\gamma = 0$, $\kappa = -3.4$; the perspective antitail observed on 2021 October 8 is populated by particles smaller than $\sim$10 $\mu$m in size; the total dust mass ejected until the latest observation on 2021 October 10 is $4.2 \times 10^7$ kg, with a maximum rate of 2.6 kg s$^{-1}$; the estimated size of the asteroid is 1.3 km.

(vii) The evolution of 2005 QN$_{173}$'s orbit and its clones forward in time for 100 kyr were provided. No differences between the nominal orbit and the orbits of the clones are visible. Small differences are seen only after 80 kyr from the beginning of integration. It was shown that Jupiter has a significant influence on the evolution of the orbit shape and its location in the Main-belt. We can also conclude that due to its influence, there is a very low dispersion of the clones of the nominal orbit.

(viii) We followed the evolution of the 2005 QN$_{173}$ orbit and the orbits of the sample of the 464 short-periodic comets, 8 of which approached the asteroid's orbit, at least once during the next 100 yr. The activity was also confirmed for two comets (64P/Swift-Gehrels and 120P/Mueller 1). We concluded that these objects are not genetically related, despite the very close distance of their orbits for a relatively long time. The possible approach and the similarity of the orbits we can consider as random.

## ACKNOWLEDGEMENTS

The authors are grateful to Dr Ludmilla Kolokolova and the anonymous referee for constructive suggestions and comments that helped us improve the manuscript. Also, the authors thank Dr Olga Voziakova for data obtained at the 2.5-m telescope. The work by OI, DT, OSh, and MH was supported by the Slovak Grant Agency for Science VEGA (Grant No. 2/0059/22, No. 2/0009/22) and by the Slovak Research and Development Agency under Contract No. APVV-19-0072. FM acknowledges financial support from grants MCIN/AEI /10.13039/501100011033 (Grant PID2021-123370OB-I00) and CEX2021-001131-S funded by MCIN/AEI 10.13039/501100011033. This work has made use of data from the Asteroid Terrestrial-impact Last Alert System (ATLAS) project. ATLAS is primarily funded to search for near-earth asteroids through NASA grants NN12AR55G, 80NSSC18K0284, and 80NSSC18K1575; byproducts of the NEO search include images and catalogs from the survey area. The ATLAS science products have been made possible through the contributions of the University of Hawaii Institute for Astronomy, the Queen's University Belfast, the Space Telescope Science Institute, and the South African Astronomical Observatory. OI and OSh acknowledge the support from the Erasmus + program of the European Union under grant No. 2020-1-CZ01-KA203-078200. OI was partially supported by the grant of the Ministry of Education and Science of Ukraine No. 0122U001560. OSh acknowledges the support from the Government Office of the Slovak Republic within NextGenerationEU programme under project No. 09I03-03-V01-00001. JL acknowledges support from the ACIISI, Consejería de Economía, Conocimiento y Empleo del Gobierno de Canarias and the European Regional Development Fund (ERDF) under grant with reference ProID2021010134, and support from the Agencia Estatal de Investigacion del Ministerio de Ciencia e Innovacion (AEI-MCINN) under grant 'Hydrated Minerals and Organic Compounds in Primitive Asteroids' with reference PID2020-120464GB-100. The work of IL was partially supported by the Ministry of Education and Science of Ukraine No. 0122U001911 and a grant of the Ministry of Education and Science of Ukraine for perspective development of a scientific direction 'Mathematical sciences and natural sciences' at Taras Shevchenko National University of Kyiv. JM was supported by the European Union's Horizon 2020 research and innovation program under grant agreement no. 757390.